# KRAFT: A Transaction-Level Dataset for Korean Apartment Sales Integrated with Contextual Indicators

Sejin Myung[1], Hyungjoon Kim[1*]

[1]Department of Computer Engineering, Changwon National University, 20, Changwondaehak-ro, Uichang-gu, Changwon-si, Gyeongsangnam-do, Republic of Korea
[*]Corresponding Author: Hyungjoon Kim (hyungjoon@changwon.ac.kr)

**Abstract**

Apartment transaction records are useful for studying housing markets, household finance, regional economics, and macro-financial transmission, but transaction data are often distributed separately from contextual socioeconomic indicators. We present KRAFT, a nationwide transaction-level dataset of South Korean apartment sales from January 2015 to December 2024. The dataset contains 5,320,379 apartment sale transactions across all 17 Sido regions and includes transaction timing, administrative location, exclusive residential area, reported transaction price, floor level, and construction year. KRAFT also provides auxiliary indicators covering macro-financial conditions, demographic structure, education infrastructure, private education expenditure, housing price indices, consumer sentiment, and economic policy uncertainty. The released files are organized as year-specific transaction files and separate auxiliary data tables to preserve the original temporal and spatial resolution of each source. KRAFT supports reproducible research on apartment price modeling, regional housing-market comparison, housing-demand analysis, and links between housing transactions and socioeconomic context.

## 1. Background & Summary

Housing markets are closely linked to household wealth, credit conditions, demographic change, local amenities, and expectations about future economic conditions. In the Republic of Korea, hereafter South Korea, apartments are a dominant form of urban housing: as of November 1, 2024, the 2024 register-based Population and Housing Census reported that 53.9% of general households resided in apartments [1]. Apartment prices and transactions are therefore relevant not only to housing studies, but also to research on household finance, urban and regional economics, demographic change, education demand, consumer sentiment, and macroeconomic policy transmission.

A large body of housing-market research has shown that residential property prices are affected by both asset-market dynamics and local fundamentals. Prior studies have examined housing-price persistence and market efficiency [2], the role of demographic demand in housing markets [3], and the relationship between housing valuations and economic fundamentals such as interest rates, income growth, and expected user costs [4]. For empirical studies of housing markets, transaction-level records are particularly valuable because they preserve information about individual transactions, locations, property characteristics, and timing. However, transaction records alone do not fully capture the socioeconomic environment in which housing decisions are made. Housing demand and prices may also be related to macro-financial conditions, demographic structure, household formation, school infrastructure, private education costs, housing price cycles, consumer sentiment, and economic policy uncertainty. These relationships have also been examined through hedonic pricing models that decompose housing prices into property and neighborhood attributes [20], studies of school-quality capitalization and residential sorting [21,22], macro-financial and household balance-sheet channels linking housing markets to credit, monetary policy, and consumption [23–25], policy-uncertainty measures [14], and research on spatially heterogeneous housing-supply constraints and high-demand urban markets [26–28]. Recent financial time-series studies have further shown that forecasting models can benefit from macroeconomic context, market-regime information, and cross-market dependency structures [29–31]. Although these studies focus on financial markets rather than apartment transactions directly, they reinforce the value of time-aligned contextual indicators for market-dynamics and forecasting research.

South Korea provides several publicly accessible official data sources relevant to housing-market analysis. The Ministry of Land, Infrastructure and Transport operates the Actual Transaction Price Disclosure System, which provides transaction-price data reported under the Act on Reporting Real Estate Transactions and related regulations [5]. The Bank of Korea's Economic Statistics System provides macroeconomic and financial indicators, including exchange-rate, interest-rate, monetary, consumer-sentiment, and housing-price-index series [6]. The Ministry of Data and Statistics and the Korean Statistical Information Service provide consumer-price, population, household, and private-education statistics [7–9]. The Korea Real Estate Board provides housing price trend statistics that are used for housing-market assessment and policy reference [10]. The Korea Education and Research Information Service and Schoolinfo provide school information disclosure data that can be used to construct local school infrastructure indicators [11]. In addition, Korean economic policy uncertainty indices provide information on policy-related uncertainty [13,14]. These sources are valuable, but they are distributed across separate institutions, spatial units, time formats, and file structures. This fragmentation can increase the preprocessing burden for researchers who need an integrated panel of housing transactions and contextual indicators.

Here, we present KRAFT (Korean Residential Apartment Fundamentals and Transactions), a nationwide transaction-level dataset of South Korean apartment sales integrated with macro-financial, demographic, educational, housing-market, consumer-sentiment, and policy-uncertainty indicators. KRAFT covers January 2015 to December 2024 and contains 5,320,379 apartment sale transactions across all 17 'Sido' regions of South Korea. Each transaction record includes the contract year-month, 'Sido' (first-level administrative region), 'Sigungu' (lower-level municipality or district), 'Beopjeongdong' (legal neighborhood), exclusive residential area in square meters, reported transaction price in 10,000 KRW, floor level, and construction year. The transaction records are provided as year-specific CSV files to support memory-efficient reuse; the full-period transaction table can be reconstructed by concatenating the annual files.

To support contextual analysis, KRAFT also includes auxiliary datasets aligned to the transaction period. The macro-financial component includes the KRW/USD exchange rate, the Bank of Korea base rate, mortgage-related lending rates, jeonse loan rates, personal credit loan rates, broad money supply, and the consumer price index. The demographic component includes population density, total population, total households, and net migration. The education component includes monthly counts of operating schools by school type at the 'Sigungu' level and private education expenditure indicators mapped to 'Sigungu' areas. The housing-market component includes the apartment housing sales price index, while the sentiment and uncertainty component includes consumer survey indices related to future economic conditions, interest-rate expectations, housing-price expectations, and the Korean economic policy uncertainty index. Monthly indicators are aligned using standardized 'YearMonth' keys in 'YYYYMM' format. Annual or lower-frequency indicators are retained at their original frequency and represented using year-end 'YearMonth' keys where appropriate.

Several auxiliary variables involve harmonization or mapping across spatial classifications. Consumer survey indices were collected from regional categories available in the original Bank of Korea data and expanded to the 17 'Sido' regions using documented mapping rules. Private education expenditure indicators were collected from categorical regional groups and mapped to 'Sigungu' areas according to documented region-type classifications. These mapped variables should therefore be interpreted as contextual indicators assigned to 'Sido' or 'Sigungu' areas, not as independently surveyed values for each individual administrative unit. Because the auxiliary indicators differ in temporal resolution, spatial resolution, and table format, KRAFT provides them as separate standardized files rather than forcing them into a single transaction-level table.

Administrative-region labels were harmonized using documented region crosswalk metadata. Name changes, administrative transfers, and inconsistencies in source-file naming conventions were treated separately to reduce region-label mismatches across institutions and years. The harmonization process included cases such as the standardization of Incheon Nam-gu/Michuhol-gu, the treatment of Gunwi-gun after its administrative transfer, and the unification of Hwaseong-si labels. Users conducting historical boundary analysis should consult the region crosswalk and account for these harmonization decisions. The dataset therefore provides standardized region labels and metadata that support aggregation to monthly, 'Sigungu'-level, 'Sido'-level, national, or custom spatial units depending on the research design.

KRAFT is intended to support reproducible research on South Korean housing markets at multiple spatial and temporal resolutions. Potential uses include apartment price modeling, housing-market forecasting, regional housing-market comparison, policy analysis, analysis of macro-financial transmission to housing markets, demographic and housing-demand studies, and studies linking education-related local amenities to housing transactions. Because the dataset preserves transaction-level observations while also providing standardized contextual indicators, users may aggregate the data flexibly according to their empirical design.

This dataset makes the following contributions:

1. We provide KRAFT, a nationwide transaction-level apartment sales dataset covering 5,320,379 apartment transactions across all 17 ‘Sido’ regions of South Korea from January 2015 to December 2024.
2. KRAFT integrates apartment transaction records with contextual indicators, including macro-financial conditions, demographic structure, educational infrastructure, private education expenditure, housing price indices, consumer sentiment, and economic policy uncertainty.
3. KRAFT reduces the preprocessing burden associated with fragmented public data sources by harmonizing temporal keys, administrative-region labels, variable names, units, metadata, and file structures across multiple publicly accessible official data sources.

The remainder of this paper is structured as follows. The Methods section describes the data collection and preprocessing procedures. The Data Records section describes the released files, variables, and metadata. The Technical Validation section presents data quality checks and validation results. The Usage Notes section discusses potential applications, merging strategies, and access considerations.

## 2. Methods

KRAFT was constructed by collecting, cleaning, harmonizing, and documenting apartment transaction records and contextual indicators from public and official data sources in South Korea. The primary transaction data were obtained from the Actual Transaction Price Disclosure System operated by the Ministry of Land, Infrastructure and Transport [5]. Auxiliary indicators were collected from official statistical and public data sources, including the Bank of Korea Economic Statistics System, the Ministry of Data and Statistics and Korean Statistical Information Service, Schoolinfo, housing price statistics, and the Korean Economic Policy Uncertainty database [6–14]. Figure 1 summarizes the overall workflow used to construct KRAFT, from source data collection to preprocessing, harmonization, component-level organization, and final dataset release.

The original source datasets were provided mainly in Korean and differed in source structure, variable naming, temporal resolution, spatial unit, and administrative-region notation. To construct a reusable dataset, we standardized variable names in English, converted temporal identifiers to a common ‘YearMonth’ key, retained Korean administrative-region values in location fields, and harmonized region labels across source datasets. In long-format files with explicit location fields, ‘Sido’, ‘Sigungu’, and ‘Beopjeongdong’ values were retained in Korean to maintain compatibility with official Korean source data. Some ‘Sido’-level auxiliary files are provided in wide format with one column per ‘Sido’ region using English romanized region names. Released variable names were standardized in English for accessibility and consistency.

The preprocessing workflow included removing canceled apartment transactions, parsing address fields into administrative units, removing exact duplicate transaction records, standardizing administrative-region labels, and aligning auxiliary indicators to their original spatial and temporal resolutions. Monthly variables were represented using ‘YearMonth’ in ‘YYYYMM’ format. Annual variables were retained at their original annual frequency and represented using year-end ‘YearMonth’ keys where appropriate. Transaction records are provided as year-specific files from 2015 to 2024, and the full-period transaction table can be reconstructed by concatenating the annual files.

Because the auxiliary indicators differ in spatial and temporal resolution, they were not forcibly merged into a single transaction-level table. Instead, KRAFT provides separate standardized data tables that can be reshaped to long format before merging. Wide-format ‘Sido’-level auxiliary files can be reshaped to long format before merging with transaction records or aggregated panels. This structure preserves transaction-level detail while allowing users to construct analysis-specific panels at transaction-level, monthly, ‘Sigungu’-level, ‘Sido’-level, national, or custom aggregation levels.

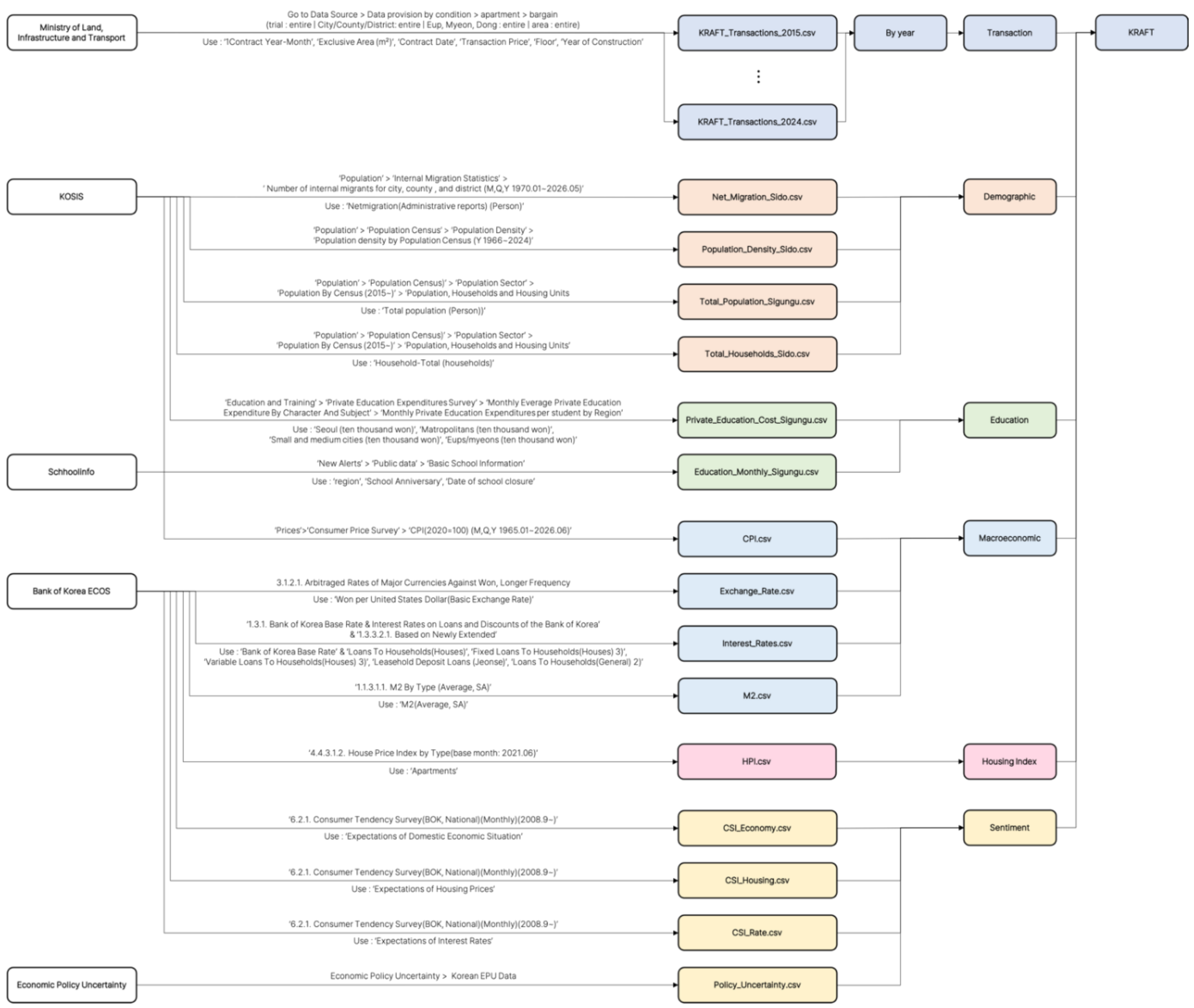

**Figure 1. Overview of the KRAFT dataset construction workflow**

*2.1. Data sources and collection scope*

The data sources for KRAFT were selected based on their relevance to apartment transactions and the socioeconomic conditions surrounding housing-market activity in South Korea. The primary transaction data were obtained from the Actual Transaction Price Disclosure System operated by the Ministry of Land, Infrastructure and Transport [5]. Apartment sales transaction records were collected for all available 'Sido', 'Sigungu', 'Beopjeongdong', and exclusive-area categories from January 2015 to December 2024. Table 1 summarizes the data sources, collection scope, temporal and spatial units, and main preprocessing steps used to construct KRAFT.

To provide contextual indicators for housing-market analysis, we collected macro-financial, housing-market, demographic, education, sentiment, and uncertainty variables from public and official data sources. Macro-financial indicators, including exchange rates, interest rates, and M2 money supply, were collected from the Bank of Korea Economic Statistics System [6]. The consumer price index was collected from official consumer price statistics provided through the Korean Statistical Information Service and the Ministry of Data and Statistics [7,8].

Housing price index data were collected through the Bank of Korea Economic Statistics System from the housing sales price index by housing type series and standardized to the 'Sido' labels used in the transaction data [6,10]. Demographic indicators, including population density, total population, total households, and net migration, were collected from the Korean Statistical Information Service and the Ministry of Data and Statistics [8].

Education-related indicators were collected from Schoolinfo, education open-data services, and official private education expenditure statistics [9,11]. School infrastructure variables were constructed from school basic information, while private education expenditure indicators were collected from official survey-based statistics. Sentiment and uncertainty indicators were collected from the Bank of Korea Consumer Survey Index series and the Korean Economic Policy Uncertainty database [12–14].

The temporal scope of KRAFT covers January 2015 to December 2024. Monthly variables were collected or converted to monthly YearMonth keys in YYYYMM format. Annual variables were retained at their original annual frequency and represented using year-end YearMonth keys where appropriate. The spatial scope covers all 17 'Sido' regions of South Korea. Transaction records were preserved at the individual transaction level, while auxiliary indicators were retained at the finest consistent spatial and temporal resolution supported by each source, including national, 'Sido', and 'Sigungu' units.

**Table 1. Source data and preprocessing summary for KRAFT**

| Component | Source | Original data | Period | Temporal unit | Spatial unit | Main preprocessing |
|---|---|---|---|---|---|---|
| Apartment transactions | Ministry of Land, Infrastructure and Transport | Apartment sale actual transaction records | Jan 2015–Dec 2024 | Transaction month | 'Sido', 'Sigungu', 'Beopjeongdong' | Removed canceled records, separated address fields, removed duplicate records, and harmonized administrative-region labels |
| Exchange rate | Bank of Korea ECOS | KRW/USD basic exchange rate | Jan 2015–Dec 2024 | Monthly | National | Converted to YearMonth format |
| Interest rates | Bank of Korea ECOS | Base rate, mortgage loan rates, fixed-rate mortgage loan rates, variable-rate mortgage loan rates, jeonse loan rates, and personal credit loan rates | Jan 2015–Dec 2024 | Monthly | National | Converted to YearMonth format |
| Money supply and consumer prices | Bank of Korea ECOS; KOSIS | M2 average seasonally adjusted series; Consumer Price Index, 2020=100 | Jan 2015–Dec 2024 | Monthly | National | Converted to YearMonth format |
| Housing price index | Bank of Korea ECOS / Korea Real Estate Board | Housing sales price index by housing type, reference month June 2021 | Jan 2015–Dec 2024 | Monthly | 'Sido' | Harmonized region labels to match transaction records |
| Demographic indicators | KOSIS / Ministry of Data and Statistics | Population density, total population, total households, and net migration | 2015–2024 | Monthly for net migration; annual or census-based for other demographic indicators | 'Sido' or 'Sigungu' | Cleaned hierarchical region names, selected the KOSIS net-migration variable, converted monthly migration records to 'YearMonth' format, and harmonized administrative-region labels |
| Education indicators | Schoolinfo; KOSIS / Ministry of Data and Statistics | School basic information; private education expenditure | 2015–2024 | Monthly for school counts; annual for private education expenditure | 'Sigungu' or mapped region-type categories | Counted operating schools using establishment and closure dates; mapped private education expenditure categories to 'Sigungu' areas |
| Sentiment and uncertainty | Bank of Korea ECOS; Economic Policy Uncertainty | Future economic outlook CSI, interest-rate outlook CSI, housing-price outlook CSI, and Korean EPU index | Jan 2015–Dec 2024 | Monthly | National or mapped 'Sido' categories | Converted date formats to 'YearMonth'; mapped CSI regional groups to 17 'Sido' regions |

*2.2. Apartment transaction data collection and preprocessing*

Apartment sales transaction records were collected from the Actual Transaction Price Disclosure System operated by the Ministry of Land, Infrastructure and Transport [5]. The source system provides reported real-estate transaction prices by transaction type, region, contract period, and property characteristics. For KRAFT, apartment sale records were collected for all available 'Sido', 'Sigungu', 'Beopjeongdong', and exclusive-area categories from January 2015 to December 2024. No additional regional or area-size filters were applied during collection.

The raw transaction records contained address information, contract timing, exclusive residential area, reported transaction price, floor, construction year, and cancellation-related information where available. The first preprocessing step was to exclude canceled transactions so that the released transaction records represent non-cancelled reported apartment sales. The address fields were then parsed into three administrative-location variables: 'Sido', 'Sigungu', and 'Beopjeongdong'. The contract year and month were converted into the standardized YearMonth key in YYYYMM format.

After temporal standardization and address parsing, exact duplicate transaction records were removed. Administrative-region labels were harmonized to reduce inconsistencies caused by name changes, boundary changes, and source-specific notation differences. This harmonization was necessary because the transaction records were later aligned with auxiliary indicators collected from multiple institutions that may use different labels or administrative hierarchies for the same region. Major region-label standardization cases are described in the administrative harmonization section.

The final transaction data preserve individual apartment sale observations rather than aggregated prices. Each transaction record includes administrative-location fields, contract year-month, exclusive residential area, reported transaction price, floor, and construction year. Some observations have a construction year later than the contract year. These records were retained because they were present in the source transaction records and may correspond to pre-completion or scheduled-completion apartment transactions. Users should consider this feature when filtering the data for specific empirical designs.

The cleaned transaction records are provided as year-specific files from 2015 to 2024 to support memory-efficient reuse. All annual transaction files follow the same schema, allowing users to reconstruct the full-period transaction table by concatenating the annual files. This structure preserves transaction-level detail while allowing users to analyze either the full ten-year period or selected subperiods according to their research design.

### *2.3. Auxiliary indicator construction*

Auxiliary indicators were constructed to provide macro-financial, housing-market, demographic, educational, consumer-sentiment, and uncertainty context for the apartment transaction records. These indicators were collected from multiple public and official sources and differ in their original temporal resolution, spatial unit, variable definition, and file structure. Therefore, KRAFT does not force all auxiliary variables into a single transaction-level table. Instead, each auxiliary component is retained at the finest consistent spatial and temporal resolution supported by its source data.

All auxiliary indicators were standardized using common temporal and regional identifiers where applicable. Monthly indicators were converted to YearMonth keys in YYYYMM format, while annual indicators were retained at their original annual frequency and represented using year-end YearMonth keys where appropriate. Region-specific indicators were harmonized to the administrative-region labels used in the transaction data, enabling users to link auxiliary indicators with transaction records or aggregated panels using ‘YearMonth’ and, where applicable, ‘Sido’ and ‘Sigungu’.

The following subsections describe the construction of macro-financial indicators, consumer sentiment and economic policy uncertainty indicators, the housing price index, demographic indicators, and education indicators.

#### *2.3.1 Macro-financial indicators*

Macro-financial indicators were constructed to capture national economic and financial conditions that may affect apartment market activity. The macro-financial component includes exchange rates, policy and lending interest rates, broad money supply, and consumer prices. Exchange-rate, interest-rate, and M2 money supply series were collected from the Bank of Korea Economic Statistics System [6], while the consumer price index was collected from official consumer price statistics provided through the Korean Statistical Information Service and the Ministry of Data and Statistics [7,8].

The exchange-rate series consists of the KRW/USD basic exchange rate. The interest-rate series include the Bank of Korea base rate, average mortgage loan rate, fixed-rate mortgage loan rate, variable-rate mortgage loan rate, jeonse loan rate, and general credit loan rate. The monetary series consists of M2 money supply, represented as the average balance seasonally adjusted series. The price series consists of the consumer price index, with 2020 as the base year.

All macro-financial indicators were collected for the January 2015 to December 2024 period and standardized to the common YearMonth key in YYYYMM format. Interest-rate variables are expressed as annual percentage rates, M2 is represented in billion KRW, and the consumer price index is retained as an index with 2020=100. Because these variables are national-level monthly indicators, they can be linked to apartment transaction records by YearMonth when users construct transaction-level or aggregated monthly panels.

*2.3.2 Consumer sentiment and economic policy uncertainty*

Consumer sentiment and economic policy uncertainty indicators were included to capture forward-looking perceptions and uncertainty that may influence housing-market activity. The sentiment component was constructed from the Bank of Korea Consumer Survey Index series [12]. Three indices were selected: the Consumer Survey Index for future economic outlook, the Consumer Survey Index for interest-rate-level expectations, and the Consumer Survey Index for housing-price expectations. These variables were collected for the January 2015 to December 2024 period.

The original Consumer Survey Index data were provided using regional groupings rather than fully harmonized 'Sido' values. The original regional CSI categories consisted of Seoul, six metropolitan cities, and other regions. To make these indicators compatible with the regional structure of the transaction and housing-market data, these three source-region groups were expanded and mapped to the 17 'Sido' regions using documented mapping rules. The time variable was converted to the common YearMonth key in YYYYMM format. The resulting sentiment indicators can be linked to transaction records or regional panels using 'YearMonth' and, where applicable, 'Sido'. For wide-format CSI files, users can reshape the 'Sido' columns to long format before merging. The economic policy uncertainty component was constructed from the Korean Economic Policy Uncertainty Index [13,14]. The original date format was converted to the common YearMonth format, and the series was restricted to the January 2015 to December 2024 period. Unlike the Consumer Survey Index variables, the Korean Economic Policy Uncertainty Index is retained as a national-level monthly uncertainty indicator. It can therefore be merged with transaction-level or aggregated housing-market data using YearMonth.

Together, the consumer sentiment and economic policy uncertainty indicators provide expectation- and uncertainty-related contextual variables for studies of apartment transactions, housing-price dynamics, and macro-financial transmission in South Korea.

*2.3.3 Housing price index*

The housing-market component was constructed using the apartment housing price index. The index was collected through the Bank of Korea Economic Statistics System from the housing sales price index by housing type series, with June 2021 as the reference month [6]. The series is based on housing price trend statistics compiled by the Korea Real Estate Board [10]. This component provides a regional housing-market indicator that can be used together with individual apartment transaction records to compare transaction-level outcomes with broader market-level price movements.

The original housing price index data were provided by region and month. During preprocessing, the time variable was converted to the common YearMonth key in YYYYMM format. Regional labels were harmonized to match the 'Sido' labels used in the transaction records. This step was necessary to ensure that the housing price index could be linked consistently with transaction records or aggregated transaction-level panels at the 'Sido' level.

The released housing price index is provided at the monthly 'Sido' level and uses June 2021 as the reference month. Because the housing price index is a monthly 'Sido'-level indicator, it can be linked to transaction records using 'YearMonth' and 'Sido', or used as a contextual variable in monthly 'Sido'-level analyses.

*2.3.4 Demographic indicators*

Demographic indicators were included to capture population-related conditions that may be associated with housing demand, household formation, and regional housing-market dynamics. The demographic component was constructed from official population, household, and internal-migration statistics provided through the Korean Statistical Information Service and the Ministry of Data and Statistics [8]. The selected indicators include population density, total population, total households, and net migration.

Population density, total population, and total households were collected from census-based population and household statistics. These source tables differ in their original spatial and temporal structures. Population density and total households were organized at the 'Sido' level, while total population was organized using 'Sido' and ‘Sigungu’ identifiers where lower-level administrative information was available. Annual demographic indicators were retained at their original annual frequency and represented using year-end ‘YearMonth’ keys where appropriate.

Net migration was collected from the ‘Number of internal migrants for city, county, and district’ series under ‘Population’ and ‘Internal Migration Statistics’ in the KOSIS Statistical Database. The study period was restricted to January 2015 through December 2024, and only the Netmigration(Administrative reports) (Person) variable was used. In the released dataset, net migration is provided as a monthly indicator and converted to the common YearMonth key in YYYYMM format.

During preprocessing, hierarchical region names were cleaned and standardized to improve consistency across years and source files. Time fields were converted to the common ‘YearMonth’ format where applicable. Administrative-region labels were harmonized to support linkage with transaction records and other auxiliary indicators. Net migration was retained as a signed variable because it represents net population inflows or outflows and can therefore take negative values. The demographic indicators can be linked to transaction records or aggregated housing-market panels using ‘YearMonth’, ‘Sido’, and, where applicable, ‘Sigungu’. Wide-format ‘Sido’-level demographic files can be reshaped to long format before merging.

#### *2.3.5 Education indicators*

Education indicators were constructed to capture local educational infrastructure and private education expenditure, which are important contextual factors in South Korean residential location choice and housing-market analysis. The education component consists of two parts: monthly school-count indicators and annual private education expenditure indicators. School-count indicators were constructed from school basic information provided through Schoolinfo and related education open-data services [11], while private education expenditure indicators were collected from official private education expenditure statistics provided through the Korean Statistical Information Service and the Ministry of Data and Statistics [9].

For the school infrastructure component, school opening and closure dates were used to determine whether each school was operating in each month during the study period. A school was counted as operating if it had opened by the corresponding month and had not closed before that month. Schools were counted through the month of closure. Monthly operating-school counts were then aggregated by ‘YearMonth’, ‘Sido’, ‘Sigungu’, and school type. The resulting variables represent the number of operating schools by school category at the ‘Sigungu’ level.

For the private education expenditure component, annual private education expenditure data were collected from survey-based official statistics. The original source data were provided by region-type categories rather than directly by apartment transaction locations or ‘Beopjeongdong’ areas. The original private education expenditure categories consisted of Seoul, metropolitan cities, small and medium-sized cities, and eup/myeon areas. These four region-type categories were mapped to ‘Sigungu’-level records using documented region-classification rules. Annual values were retained at their original annual frequency and represented using year-end ‘YearMonth’ keys.

Administrative-region labels in both education components were standardized to support linkage with transaction records and other auxiliary indicators. Special administrative cases, including region-boundary and region-type treatment, were handled consistently with the administrative harmonization rules used in the rest of the dataset. The education indicators can be linked to apartment transaction records or aggregated housing-market panels using ‘YearMonth’, ‘Sido’, and ‘Sigungu’, depending on the spatial and temporal resolution required by the analysis.

### *2.4. Administrative harmonization and dataset organization*

Administrative-region harmonization was performed to improve longitudinal consistency and cross-source compatibility. The source datasets used different administrative notations depending on the institution, year, and spatial resolution. In addition, administrative name changes, boundary transfers, and hierarchical differences could create inconsistencies when transaction records were combined with auxiliary indicators. To reduce these inconsistencies, region labels were standardized across transaction, demographic, education, housing-index, and sentiment-related files.

The released dataset retains Korean administrative-region values in long-format location fields such as 'Sido', 'Sigungu', and 'Beopjeongdong'. This design was chosen to preserve compatibility with the original Korean public data sources and to avoid ambiguity in official administrative names. At the same time, released variable names were standardized in English to improve accessibility for international users. Major administrative-region standardization cases include the harmonization of Incheon Nam-gu and Michuhol-gu 'Sigungu' labels, the unification of Hwaseong-si 'Sigungu' labels, the time-varying 'Sido' treatment of Gunwi-gun, and consistent 'Sigungu' labeling for Sejong Special Self-Governing City.

Gunwi-gun was handled according to the temporal resolution of each released data component. Monthly records follow the administrative affiliation applicable to each observation month, so records before the administrative transfer are retained under Gyeongsangbuk-do where applicable, while records after the transfer are retained under Daegu Metropolitan City. Annual records follow the administrative classification adopted for each annual source table. In the released private education expenditure data, Gunwi-gun is classified under Gyeongsangbuk-do through 2023 and under Daegu Metropolitan City from 2024.

KRAFT was organized as component-level data tables rather than as a single fully merged transaction-level panel. This structure was adopted because the transaction records and auxiliary indicators differ in their spatial and temporal resolutions. Transaction records are provided at the individual transaction level, while auxiliary indicators may be national, 'Sido'-level, 'Sigungu'-level, monthly, or annual. Providing these components separately preserves the original resolution of each source and allows users to construct analysis-specific panels according to their research design.

The cleaned apartment transaction records are released as year-specific files from 2015 to 2024. The full-period transaction table can be reconstructed by concatenating the annual files because all annual transaction files follow the same schema. Auxiliary indicators are released as separate standardized data tables and can be linked to transaction records or aggregated transaction panels using 'YearMonth' and, where applicable, 'Sido' and 'Sigungu', depending on the spatial and temporal resolution and table format of each file. Wide-format 'Sido'-level files can be reshaped to long format when users need an explicit 'Sido' merge key.

To support transparent reuse, the released dataset is accompanied by metadata and documentation describing variable definitions, source information, administrative-region harmonization rules, preprocessing procedures, and validation results. This organization allows users to trace the source and preprocessing history of each data component while retaining flexibility in how the transaction records and contextual indicators are combined.

## 3. Data Records

KRAFT is released as a structured collection of CSV data files and accompanying documentation files. The released data are organized into two broad groups: data component files and metadata/documentation files. The data component files include year-specific apartment transaction records and auxiliary indicators covering macro-financial conditions, housing price indices, demographic indicators, education indicators, consumer sentiment, and economic policy uncertainty. The metadata and documentation files provide variable definitions, source information, administrative-region harmonization rules, preprocessing summaries, validation results, license information, and citation metadata.

The apartment transaction records are provided as annual files from 2015 to 2024. All annual transaction files follow the same schema, allowing users to reconstruct the full transaction-level dataset by concatenating the annual files. Auxiliary indicators are provided as separate component-level files rather than being merged into a single transaction-level panel. This release structure preserves the original spatial and temporal resolution of each auxiliary indicator and allows users to construct analysis-specific datasets using YearMonth and, where applicable, Sido and Sigungu. Wide-format Sido-level files can be reshaped to long format before being linked to transaction records or aggregated panels.

All released data files use English variable names, while Korean administrative-region values are retained in location fields to preserve compatibility with the original Korean public data sources. Detailed variable-level descriptions and source-level metadata are provided in the accompanying documentation files. The following subsections describe the released data components and the metadata files included in the KRAFT release.

### *3.1 Released data components*

The released KRAFT data components consist of apartment transaction records and auxiliary indicator files. Transaction records are provided as year-specific CSV files, while auxiliary indicators are organized by data category according to their original spatial and temporal resolutions. Table 2 summarizes the main released data components, observation units, temporal units, spatial units, and merge keys.

**Table 2. Released data components and merge keys in KRAFT**

| Component | Released files | Observation unit | Temporal unit | Spatial unit | Main keys |
|---|---|---|---|---|---|
| Apartment transactions | KRAFT_Transactions_YYYY.csv | Individual apartment sale transaction | Transaction month | 'Sido', 'Sigungu', 'Beopjeongdong' | 'YearMonth', 'Sido', 'Sigungu', 'Beopjeongdong' |
| Macro-financial indicators | CPI.csv, M2.csv, Exchange_Rate.csv, Interest_Rates.csv | Monthly indicator | Monthly | National | 'YearMonth' |
| Housing price index | HPI.csv | Regional housing price index | Monthly | 'Sido' | 'YearMonth', 'Sido' |
| Demographic indicators | Net_Migration_Sido.csv, Population_Density_Sido.csv, Total_Households_Sido.csv, Total_Population_Sigungu.csv | Regional demographic indicator | Monthly or annual | 'Sido' or 'Sigungu' | 'YearMonth'; 'YearMonth' and 'Sido' after reshaping where applicable; 'YearMonth', 'Sido', and 'Sigungu' for long-format 'Sigungu' files |
| Education indicators | Education_Monthly_Sigungu.csv, Private_Education_Cost_Sigungu.csv | Regional education indicator | Monthly or annual | 'Sigungu' | 'YearMonth', 'Sido' and normalized parent-city 'Sigungu' where required |
| Consumer sentiment | CSI_Economy.csv, CSI_Housing.csv, CSI_Rate.csv | Regionalized sentiment indicator | Monthly | mapped 'Sido' | 'YearMonth', 'Sido' after reshaping where applicable |
| Economic policy uncertainty | Policy_Uncertainty.csv | National uncertainty indicator | Monthly | National | 'YearMonth' |

The apartment transaction component contains 5,320,379 individual apartment sale records from January 2015 to December 2024. The transaction records are released as ten annual files, one for each year from 2015 to 2024. Each annual file follows the same schema and includes administrative-location fields, transaction timing, exclusive residential area, reported transaction price, floor, and construction year. The full transaction table can be reconstructed by concatenating the annual files.

The macro-financial component contains national-level monthly indicators. These files include the consumer price index, M2 money supply, KRW/USD exchange rate, and interest-rate variables. The interest-rate file includes the Bank of Korea base rate, mortgage-related lending rates, jeonse loan rate, and general credit loan rate. These indicators can be linked to transaction records or monthly aggregated panels using 'YearMonth'.

The housing price index component contains monthly apartment housing price index values at the 'Sido' level. This component provides regional housing-market context and can be linked to transaction records or aggregated regional panels using 'YearMonth' and 'Sido'.

The demographic component includes population density, total households, total population, and net migration. These indicators are released at the spatial and temporal resolutions supported by the source data. Some files are organized at the 'Sido' level, while others use both 'Sido' and 'Sigungu' identifiers. Monthly indicators are indexed by monthly 'YearMonth' values, and annual indicators are represented using year-end 'YearMonth' keys where appropriate.

The education component includes monthly school-count indicators and annual private education expenditure indicators. The school-count file provides the number of operating schools by school type at the 'Sigungu' level. The private education expenditure file provides annual expenditure indicators mapped to 'Sigungu'-level records according to the documented region-classification rules. These files can be linked using 'YearMonth', 'Sido', and 'Sigungu', depending on the intended spatial and temporal resolution of analysis.

The consumer sentiment and economic policy uncertainty component includes three Consumer Survey Index files and one Korean Economic Policy Uncertainty Index file. The Consumer Survey Index files provide regionalized indicators for future economic outlook, housing-price expectations, and interest-rate expectations. The economic policy uncertainty file provides a national-level monthly uncertainty indicator. These indicators can be used as expectation- and uncertainty-related contextual variables in transaction-level or aggregated housing-market analyses.

### *3.2 Metadata and documentation files*

KRAFT includes metadata and documentation files to support transparent reuse, source traceability, administrative-region interpretation, and reproducible data integration. These files describe the released variables, source datasets, preprocessing decisions, validation results, license conditions, and citation information. They are provided together with the data component files in the root directory of the released dataset package.

'Variable_Dictionary.csv' provides variable-level metadata for the released data files. It includes variable names, descriptions, data types, units, spatial and temporal levels, and notes on variable interpretation where applicable. This file is intended to help users understand the meaning and expected use of each variable before merging or analyzing the data.

'Source_Metadata.csv' summarizes the source information for each released data component. It records the source institution, source dataset or statistical series, collection period, temporal resolution, spatial resolution, source access information, and main preprocessing steps. This file allows users to trace each released component back to its original public data source and understand how it was standardized for inclusion in KRAFT.

'Region_Mapping.csv' documents administrative-region harmonization rules used during dataset construction. It includes region-label standardization cases, historical or administrative-boundary-related treatments, and join-related decisions for regions requiring special handling. This file is particularly important for users who merge KRAFT with other Korean administrative or regional datasets.

'Preprocessing_Summary.md' provides a narrative description of the main preprocessing procedures applied to the transaction records and auxiliary indicators. It summarizes steps such as canceled-transaction removal, duplicate-record removal, temporal standardization, administrative-region harmonization, source-specific processing, and dataset organization. This document complements the Methods section by providing a reusable reference for users of the dataset package.

'Validation_Report.csv' summarizes the technical validation checks performed on the released files. It includes checks for file completeness, schema consistency, temporal coverage, regional coverage, missing values, duplicate records, plausible value ranges, and metadata consistency. This file provides a machine-readable summary of the validation procedures and results.

'README.md' provides an overview of the dataset, directory structure, released files, recommended usage, merge keys, license information, and citation instructions. 'LICENSE' describes the license applied to the KRAFT dataset compilation and clarifies that the original source data remain subject to the terms of the respective source institutions. 'CITATION.cff' provides structured citation metadata for users who cite the dataset in academic publications or software environments.

Together, these metadata and documentation files are intended to make KRAFT reusable beyond the original construction context. They allow users to identify relevant files, understand variable definitions, trace source data, interpret administrative-region harmonization rules, verify validation outcomes, and cite the dataset consistently.

## 4. Technical Validation

The KRAFT dataset was technically validated to assess the completeness, consistency, plausibility, and reusability of the released data files. The validation process focused on both data integrity and descriptive validation. Data integrity checks were performed to confirm that the released files were complete, readable, consistently structured, and aligned with the intended temporal and regional coverage. Descriptive validation was performed to summarize the distributions of key transaction-level variables and selected auxiliary indicators, following the common practice of reporting descriptive statistics in dataset publications.

The validation covered the apartment transaction records, macro-financial indicators, housing price indices, demographic indicators, education indicators, consumer sentiment indicators, and economic policy uncertainty series. For the transaction component, we checked annual file completeness, schema consistency across yearly files, temporal coverage from January 2015 to December 2024, coverage of all 17 regions, missing administrative identifiers, duplicate records, and plausible value ranges for key numerical variables. For the auxiliary components, we checked temporal coverage, regional coverage where applicable, duplicate regional-time records, missing values, and plausible ranges according to the definition of each variable.

In addition to integrity checks, descriptive statistics were computed for the main transaction-level variables and selected auxiliary indicators. These summaries were used to verify that the released variables had interpretable ranges and distributions. For transaction records, the validation focused on variables such as transaction price, exclusive residential area, floor, and construction year. For auxiliary indicators, representative variables were selected from the macro-financial, housing-market, demographic, education, sentiment, and uncertainty components.

The following subsections describe the validation procedures and results. Section 4.1 summarizes file completeness, schema consistency, and coverage validation. Sections 4.2 and 4.3 present descriptive validation of transaction records and auxiliary indicators, respectively. Section 4.4 reports distributional and regional coverage checks. Section 4.5 summarizes metadata consistency checks and the validation outputs included in the released dataset package.

### *4.1 File completeness, schema consistency, and coverage validation*

File-level and schema-level validation was performed to confirm that the released KRAFT package contained the expected data components and that the files could be reused consistently. All released CSV files were checked for readability, file completeness, column availability, and compatibility with the documented file structure. The ten annual transaction files from 2015 to 2024 were also checked to ensure that they followed the same schema, so that the full transaction-level dataset could be reconstructed by concatenating the annual files.

**Table 3. Summary of data integrity validation results**

| Validation item | Result |
| --- | --- |
| Annual transaction files | 10 files, 2015–2024 |
| Total transaction records | 5,320,379 |
| Transaction period | 201501–202412 |
| Monthly transaction periods | 120 months |
| Sido coverage in transaction records | 17 first-level administrative regions |
| Transaction file schema consistency | Passed |
| Missing Sido, Sigungu, or Beopjeongdong values | None detected |
| Exact duplicate transaction rows | None detected |
| Non-positive transaction prices | None detected |
| Non-positive exclusive residential areas | None detected |
| Macro-financial monthly coverage | 201501–202412 |
| Housing price index coverage | 201501–202412, 17 ‘Sido’ regions |
| Demographic indicator coverage | Checked according to source-specific spatial and temporal units |
| Education monthly coverage | 201501–202412, regional coverage checked |
| Consumer sentiment coverage | 201501–202412, regionalized Sido-level coverage checked |
| Economic policy uncertainty coverage | 201501–202412 |
| Metadata consistency | Released files and variables documented |

The apartment transaction component was validated for temporal and regional completeness. The released transaction files contain 5,320,379 apartment sale records covering the period from January 2015 to December 2024. All 120 monthly periods in the study window were present, and all 17 first-level administrative regions of South Korea were represented. Required administrative-location fields, including Sido, Sigungu, and Beopjeongdong, were checked for missing values. The transaction records were also checked for exact duplicate rows and non-positive values in key numerical variables such as transaction price and exclusive residential area.

Auxiliary indicator files were validated according to their expected temporal and spatial resolutions. National-level monthly files were checked for complete monthly coverage from January 2015 to December 2024. Region-level files were checked for the expected 'Sido' or 'Sigungu' identifiers, depending on the spatial resolution of the corresponding component. Monthly regional files were checked for duplicated region-month records, and annual files were checked for duplicated region-year records. Missing values and implausible value ranges were also inspected according to the definition of each variable.

Table 3 summarizes the main data integrity validation results. More detailed validation outputs are provided in Validation_Report.csv, which is included in the released dataset package.

### *4.2 Descriptive validation of transaction records*

Descriptive validation was conducted for the main transaction-level variables to examine whether the released apartment transaction records have interpretable ranges and distributions. Following prior dataset studies that reported descriptive statistics, variable ranges, and distributional summaries as part of dataset description or validation [15–18], we summarized the core numerical variables using valid observations, mean, standard deviation, minimum, first quartile, median, third quartile, and maximum values. In addition, because housing-market datasets commonly include transaction prices and property characteristics as key variables [19], we also inspected the relationship between transaction price and exclusive residential area through a validation-purpose price-per-square-meter variable.

For each numerical variable (x), the number of valid observations was calculated as Eq (1).

$$N_x = \sum_{i=1}^{N} I(x_i \neq \text{missing}) \tag{1}$$

where $(I(\cdot))$ is an indicator function. The sample mean and standard deviation were computed as Eq (2) and Eq (3).

$$\bar{x} = \frac{1}{N_x}\sum_{i=1}^{N_x} x_i \tag{2}$$

$$s_x = \sqrt{\frac{1}{N_x - 1}\sum_{i=1}^{N_x}(x_i - \bar{x})^2} \tag{3}$$

Quartiles were calculated as the 25th, 50th, and 75th percentiles of each variable distribution as described in Eq (4):

$$Q_{1(x)} = \text{Quantile}_{0.25}(x), \qquad Q_{2(x)} = \text{Quantile}_{0.50}(x), \qquad Q_{3(x)} = \text{Quantile}_{0.75}(x) \tag{4}$$

Here, $Q_2(x)$ corresponds to the median. For validation purposes, we additionally calculated transaction price per square meter as Eq (5).

$$\text{Price_per_sqm_10k_KRW}_i = \frac{\text{Transaction_Price_10k_KRW}_i}{\text{Exclusive_Area_sqm}_i} \quad (5)$$

This derived value was used only to inspect the plausibility of transaction prices relative to residential area and is not required as a released variable.

Table 4 reports the descriptive statistics of the core transaction-level variables. The transaction price variable had a mean of 33,703.180 and a median of 25,400.000 in units of 10,000 KRW, with values ranging from 400.000 to 2,500,000.000. The mean was larger than the median, indicating a right-skewed distribution that is consistent with heterogeneous apartment prices across regions, unit sizes, and market segments. Exclusive residential area had a mean of 75.158 square meters and a median of 76.000 square meters, with an interquartile range from 59.730 to 84.960 square meters.

Floor values ranged from -4.000 to 83.000, with a median of 8.000. Negative floor values were present in 281 records and correspond to basement or below-ground floor notation in the source data rather than preprocessing errors. Construction year ranged from 1,961 to 2,025, with a median of 2,001. A total of 225 records, corresponding to 0.0042% of transaction records, had construction years later than the contract year. These records were retained because they were present in the source transaction records and may correspond to pre-completion or scheduled-completion apartment transactions. They were treated as source-consistent records rather than automatically removed. The validation-purpose price-per-square-meter variable had a mean of 436.899 and a median of 341.373 in units of 10,000 KRW per square meter. Its maximum value was 9,126.006, reflecting the right-skewed nature of apartment prices. Because apartment transaction prices are expected to exhibit right-skewed distributions due to regional, size, and quality differences, extreme high values were inspected as part of distributional validation rather than automatically removed. Overall, the descriptive statistics indicate that the transaction-level variables have plausible ranges and are suitable for reuse in transaction-level or aggregated housing-market analyses.

**Table 4. Descriptive statistics of transaction-level variables.**

| Variable | Mean | Std. | Min | Q1 | Median | Q3 | Max |
|---|---|---|---|---|---|---|---|
| Transaction_Price_10k_KRW | 33,703.180 | 32,687.463 | 400.000 | 15,400.000 | 25,400.000 | 41,000.000 | 2,500,000.000 |
| Exclusive_Area_sqm | 75.158 | 25.850 | 9.260 | 59.730 | 76.000 | 84.960 | 424.320 |
| Floor | 9.254 | 6.398 | -4.000 | 4.000 | 8.000 | 13.000 | 83.000 |
| Construction_Year | 2,001.936 | 9.665 | 1,961.000 | 1,995.000 | 2,001.000 | 2,009.000 | 2,025.000 |
| Price_per_sqm_10k_KRW | 436.899 | 350.986 | 7.583 | 232.480 | 341.373 | 516.375 | 9,126.006 |

*4.3 Descriptive validation of auxiliary indicators*

Descriptive validation was also conducted for the auxiliary indicators to verify that the contextual variables included in KRAFT have interpretable ranges and are consistent with their expected measurement units and spatial-temporal resolutions. Because the auxiliary files were released at different levels of aggregation, descriptive statistics were computed at the observation level of each file. National monthly indicators were summarized over monthly observations, Sido-level indicators were summarized over region-month or region-year observations, and Sigungu-level indicators were summarized over local-government-time observations.

For wide regional files, such as Sido-level demographic or consumer sentiment files, the regional columns were reshaped into a long region-time format before computing descriptive statistics. For education indicators, monthly school counts were inspected by school type, and a validation-purpose total school-count variable was computed as Eq (6).

$$Total_School_Count_{k,r,t} = \sum_{k \in K} School_Count_{k,r,t} \quad (6)$$

where r denotes a Sigungu-level administrative unit, t denotes month, and K is the set of school categories included in the released education file. This derived value was used only to validate the overall range of local educational infrastructure and is not required as a released variable.

Table 5 reports descriptive statistics for selected auxiliary indicators. The macro-financial variables covered the full monthly period and showed plausible ranges. The Bank of Korea base rate ranged from 0.500% to 3.500%, while the average mortgage loan rate ranged from 2.390% to 4.820%. The KRW/USD exchange rate ranged from 1,066.700 to 1,434.420, M2 money supply ranged from 1,949,424.500 to 3,900,070.900 billion KRW, and the consumer price index ranged from 94.587 to 114.910.

**Table 5. Descriptive statistics of selected auxiliary indicators**

| Component | Var | Observation level | N | Mean | Std. | Min | Q1 | Median | Q3 | Max |
|---|---|---|---|---|---|---|---|---|---|---|
| MF | Base_Rate_Percent | NM | 120 | 1.773 | 0.982 | 0.500 | 1.250 | 1.500 | 2.062 | 3.500 |
| MF | Mortgage_Rate_Avg_Percent | NM | 120 | 3.328 | 0.647 | 2.390 | 2.810 | 3.215 | 3.857 | 4.820 |
| MF | Exchange_Rate_KRW_per_USD | NM | 120 | 1,197.295 | 92.863 | 1,066.700 | 1,125.280 | 1,175.465 | 1,270.095 | 1,434.420 |
| MF | M2_Money_Supply_Billion_KRW | NM | 120 | 2,890,115.638 | 630,111.359 | 1,949,424.500 | 2,335,533.675 | 2,730,220.950 | 3,555,841.425 | 3,900,070.900 |
| MF | CPI_2020_100 | NM | 120 | 102.282 | 6.392 | 94.587 | 97.560 | 99.756 | 108.315 | 114.910 |
| HPI | HPI_2021_06_100 | SDM | 2,040 | 93.520 | 10.692 | 64.600 | 88.100 | 95.300 | 101.000 | 115.300 |
| DG | Population_Density | SDY | 170 | 2,121.685 | 3,676.980 | 90.200 | 221.775 | 744.300 | 2,771.275 | 16,364.000 |
| DG | Total_Households | SDY | 170 | 1,250,429.935 | 1,309,122.572 | 76,419.000 | 617,758.000 | 794,720.000 | 1,240,971.750 | 5,827,597.000 |
| DG | Total_Population | SGY | 2,496 | 204,072.595 | 161,524.064 | 8,169.000 | 56,594.500 | 174,825.000 | 314,599.250 | 1,004,079.000 |
| DG | Net_Migration | SDM | 2,040 | 0.000 | 3,417.636 | -17,393.000 | -1,001.000 | -242.000 | 608.250 | 21,855.000 |
| EDU | Total_School_Count | SGM | 27,480 | 53.512 | 36.133 | 6.000 | 29.000 | 44.000 | 64.000 | 236.000 |
| EDU | Private_Education_Expense_10k_KRW | SGY | 2,520 | 31.218 | 10.508 | 16.000 | 24.700 | 29.000 | 39.100 | 67.300 |
| CS | CSI_Economy | SDM | 2,040 | 81.634 | 12.960 | 50.000 | 74.000 | 81.000 | 90.000 | 114.000 |
| CS | CSI_Housing | SDM | 2,040 | 105.975 | 15.419 | 60.000 | 98.000 | 107.000 | 117.000 | 137.000 |
| CS | CSI_Rate | SDM | 2,040 | 112.536 | 18.521 | 70.000 | 96.000 | 113.000 | 126.000 | 153.000 |
| EPU | Economic_Policy_Uncertainty_Index | NM | 120 | 204.733 | 102.349 | 71.157 | 137.784 | 188.652 | 249.604 | 872.877 |

Where MF, HPI, DG, EDU, CS, EPU denote Macro-financial, Housing price index, Demographic, Education, Consumer sentiment, and Economic policy uncertainty, respectively. In Observation level column, NM, SDM, SDY, SGM, SGY denote National-month, Sido-month, Sido-year, Sigungu-month, and Sigungu-year, respectively.

The housing price index had a median of 95.300 and ranged from 64.600 to 115.300 across Sido-month observations. Demographic indicators also showed plausible ranges across regional observations. Population density ranged from 90.200 to 16,364.000, reflecting large differences between metropolitan and non-metropolitan regions. Total population at the Sigungu-year level ranged from 8,169.000 to 1,004,079.000. Net migration ranged from -17,393.000 to 21,855.000; negative values were retained because net migration represents net inflows or outflows and can therefore take negative values by definition.

Education indicators were checked to confirm that school-count and private-education variables had non-negative and interpretable values. The validation-purpose total school count ranged from 6.000 to 236.000 at the Sigungu-month level. Private education expenditure ranged from 16.000 to 67.300 in units of 10,000 KRW. Consumer sentiment indicators showed ranges consistent with index-type variables, with future economic outlook CSI ranging from 50.000 to 114.000, housing-price expectation CSI ranging from 60.000 to 137.000, and interest-rate expectation CSI ranging from 70.000 to 153.000. The Korean Economic Policy Uncertainty Index ranged from 71.157 to 872.877 over the monthly period.

Overall, the descriptive statistics indicate that the auxiliary indicators have plausible ranges, preserve their intended spatial and temporal resolutions, and can be reused as contextual variables for transaction-level or aggregated housing-market analyses. Full descriptive statistics for all auxiliary variables are provided in the accompanying KRAFT_Auxiliary_Descriptive_Statistics_Full.csv output generated by the validation script.

*4.4 Distributional and regional coverage checks*

Distributional and regional coverage checks were performed to complement the descriptive statistics reported in Sections 4.2 and 4.3. These checks focused on whether transaction records were continuously available over the study period, whether all first-level administrative regions were represented, and whether the validation-purpose price-per-square-meter variable showed the expected right-skewed distribution of housing transaction prices.

First, annual transaction counts were computed from the year-specific transaction files. Figure 2 shows the annual number of apartment transactions from 2015 to 2024. The annual counts confirm continuous transaction coverage across the ten-year study period. The number of annual transactions ranged from 254,273 in 2022 to 799,414 in 2020. The total number of records across the annual files was 5,320,379, which matches the total transaction count used in the integrity and descriptive validation checks.

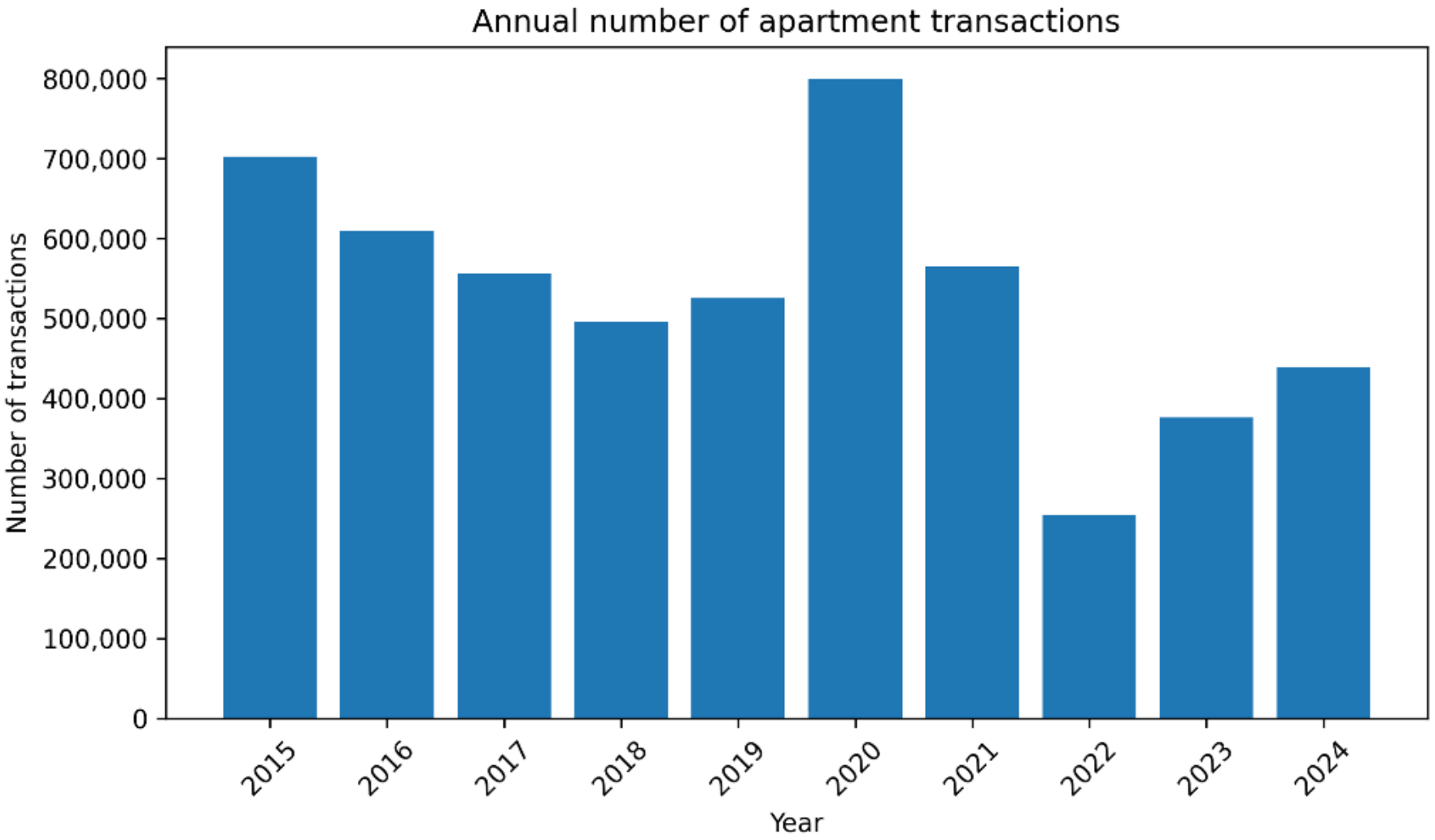

**Figure 2. Annual number of apartment transactions from 2015 to 2024**

Second, regional coverage was examined at the Sido level. Figure 3 shows the number of transaction records by Sido region. All 17 Sido regions were represented in the transaction records. The largest number of transactions was observed in Gyeonggi, with 1,461,211 records, while the smallest number was observed in Jeju, with 27,425 records. The five regions with the largest transaction counts were Gyeonggi, Seoul, Busan, Gyeongnam, and Incheon. These counts were used as regional coverage diagnostics rather than as substantive housing-market results.

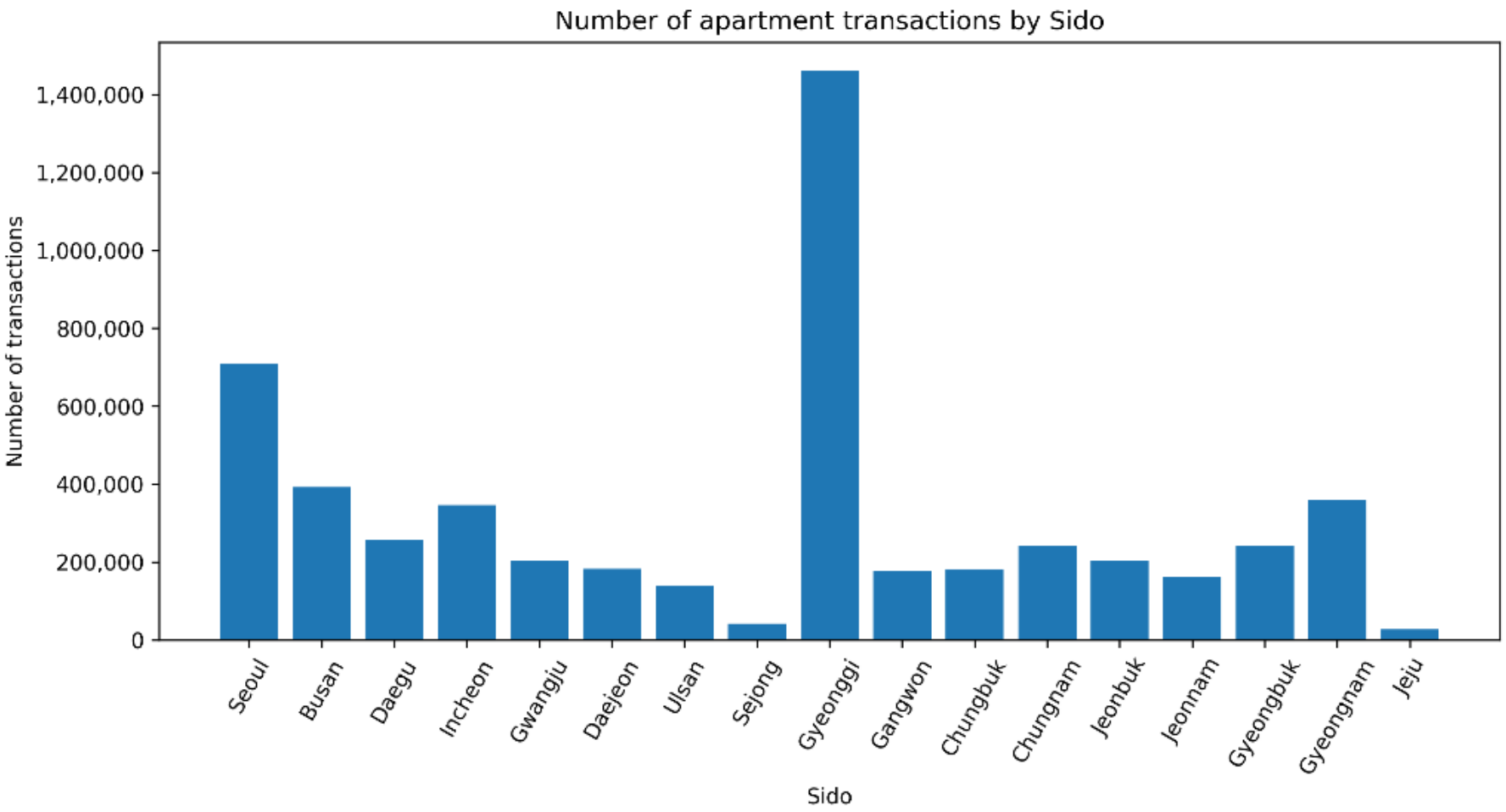

**Figure 3. Number of apartment transactions by Sido**

To further verify temporal-regional completeness, Sido-month coverage was checked for all 17 regions and 120 monthly periods. A total of 2,040 of 2,040 Sido-month combinations contained at least one transaction record. The number of transactions per Sido-month ranged from 110 to 34,237. The number of Sido-month combinations without any transaction record was 0. This result confirms that the released transaction data provide continuous regional coverage at the Sido-month level.

Finally, the distribution of the validation-purpose transaction price per square meter was inspected. Figure 4 shows the distribution of the log-transformed price-per-square-meter variable. The untransformed price per square meter had a median of 341.373 and a 99th percentile of 1,852.225 in units of 10,000 KRW per square meter, with a maximum of 9,126.006. The distribution was right-skewed, as expected for housing-market transaction prices that vary by region, unit size, construction quality, and local market conditions. These high-value observations were therefore treated as distributional features of the transaction data rather than automatically removed.

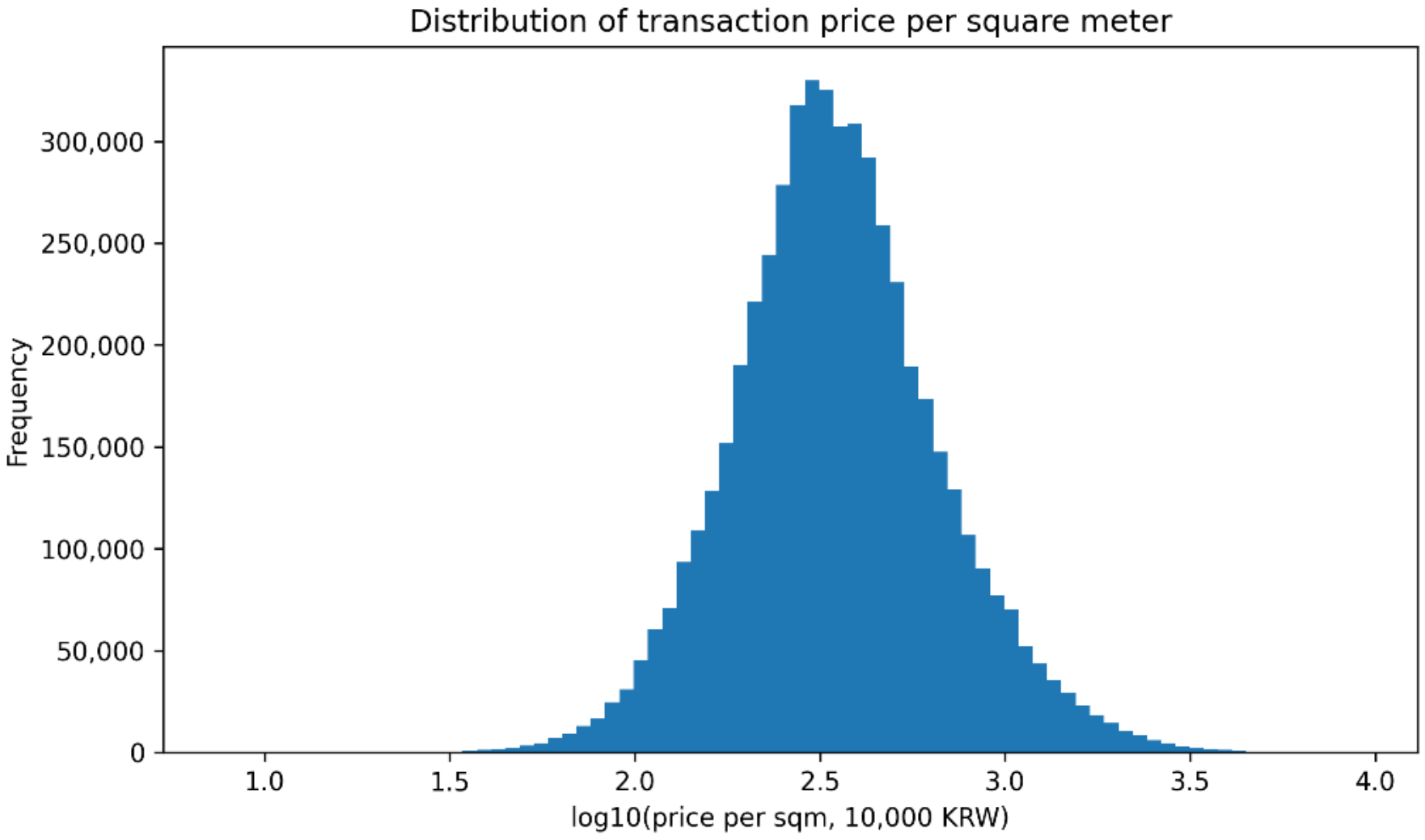

**Figure 4. Distribution of validation-purpose transaction price per square meter**

Together, the annual, regional, Sido-month, and price-distribution checks support the completeness and plausibility of the KRAFT transaction records. The generated figures and count tables are provided together with the validation code to support reproducibility.

*4.5 Metadata consistency and validation outputs*

Metadata consistency checks were performed to verify that the released data files were accompanied by sufficient documentation for transparent reuse. The checks focused on documentation-file availability, source-level metadata coverage, variable-level metadata coverage, administrative-region mapping documentation, and validation-output consistency.

The documentation files were checked to confirm that the dataset package included all expected reuse materials, including README.md, LICENSE, CITATION.cff, Variable_Dictionary.csv, Source_Metadata.csv, Region_Mapping.csv, Validation_Report.csv, and Preprocessing_Summary.md. All expected documentation files were present in the final release package. This documentation set allows users to identify the dataset structure, licensing conditions, citation information, variable definitions, source information, preprocessing decisions, administrative harmonization rules, and validation results.

Source-level consistency was assessed by comparing the released data files with 'Source_Metadata.csv'. A total of 25 of 25 released data files were documented in the source metadata file. The transaction records were documented using an annual-file template, while the auxiliary indicators were documented at the component-file level. The source metadata file records the source institution, source dataset or statistical series, collection period, temporal and spatial resolution, released variables, preprocessing summary, and license or attribution notes for each data component.

Variable-level consistency was assessed by comparing the columns in the released data files with 'Variable_Dictionary.csv'. A total of 224 of 224 file-column pairs were documented in the variable dictionary. The variable dictionary provides variable names, descriptions, units, data types, temporal resolution, spatial resolution, source information, and original variable names where applicable. In addition, 152 of 152 released-variable entries listed in 'Source_Metadata.csv' were confirmed to exist in the matched data files.

Administrative-region documentation was checked using 'Region_Mapping.csv'. This file contained 7 documented harmonization rows covering administrative-region standardization cases and join-related decisions. These records support interpretation of historical or source-specific region labels and help users reproduce joins between transaction records and auxiliary indicators.

Finally, the validation outputs were reviewed for consistency. 'Validation_Report.csv' contained 19 validation checks, all of which were marked as passing in the inspected package. The generated validation scripts and derived validation outputs provide additional reproducibility for the descriptive, distributional, regional coverage, and metadata consistency checks reported in Sections 4.2–4.5. Together, these metadata and validation outputs support the transparency, traceability, and reuse of KRAFT.

## 5. Usage Notes

KRAFT is released as component-level files rather than as a single fully merged transaction-level panel. Users should first determine the spatial and temporal resolution required by their analysis before linking auxiliary indicators to transaction records.

National monthly indicators, including exchange rates, interest rates, M2, CPI, and the Korean Economic Policy Uncertainty Index, can be linked using YearMonth. Sido-level monthly indicators, including the housing price index and regionalized Consumer Survey Index files, can be linked using YearMonth and Sido after reshaping wide-format files to long format where necessary. Sigungu-level indicators, including school-count and private education expenditure files, can be linked using YearMonth, Sido, and Sigungu where applicable.

Users should note that Consumer Survey Index variables and private education expenditure variables are mapped contextual indicators. The Consumer Survey Index variables were expanded from broader source-region groups to the 17 Sido regions, and private education expenditure variables were mapped from region-type categories to Sigungu-level records. These variables should not be interpreted as independently surveyed values for every Sido or Sigungu unit.

Annual or lower-frequency indicators represented using year-end YearMonth keys should not be interpreted as observations specific to December. The year-end key is a temporal indexing convention used to align annual indicators with monthly files. Users constructing monthly panels should decide whether to assign annual values to all months of the corresponding year, keep them as year-level variables, or use alternative temporal alignment rules depending on their research design.

Users conducting historical boundary analysis should consult Region_Mapping.csv. Gunwi-gun is handled according to the temporal resolution of each data component: monthly records follow the administrative affiliation applicable to each observation month, while annual records use year-end administrative classification. Users requiring constant-boundary panels should construct additional boundary-adjusted versions using the region-mapping metadata.

Transaction prices are expressed in units of 10,000 KRW, and exclusive residential area is expressed in square meters. Negative floor values are retained where they appear in the source records and indicate below-ground floor notation. Net migration is a signed variable; negative values indicate net outflows and should not be treated as missing values or errors.

Transaction records may preserve general-gu-level labels such as 'Yongin-si Sujigu,' whereas the education indicators are aggregated at the corresponding parent-city level, such as 'Yongin-si'. Before merging these files, users should construct a separate parent-city join key while retaining the original Sigungu value. Reusable Python code and a complete merge example are provided in the accompanying code repository.

For annual private education expenditure indicators represented using year-end YearMonth keys, users should align records by calendar year rather than interpreting YYYY12 as a December-specific observation.

### Data Availability

The KRAFT dataset described in this Data Descriptor is publicly available through Zenodo at https://doi.org/10.5281/zenodo.21320079 [32]. The repository contains annual apartment transaction files for 2015–2024, contextual indicator files, variable and source metadata, administrative-region mapping information, preprocessing documentation, and validation results.

### Code Availability

The Python code used for technical validation, figure generation, metadata consistency checks, and education-indicator merging is publicly available at https://github.com/HJCVLab/KRAFT_Validation.

### Author Contributions

Sejin Myung – Conceptualization, Data curation, Methodology, Software

Hyungjoon Kim – Validation, Visualization, Writing – original draft, Writing – review and editing, Supervision

### Competing Interests

The authors declare no competing interests.

### Acknowledgements

This research was supported by Changwon National University 2025-2026.

**References**

[1] Ministry of Data and Statistics. 2024 Population and Housing Census (Register-based Census). Press release, 29 July 2025. Ministry of Data and Statistics, Republic of Korea. https://mods.go.kr/board.es?act=view&bid=11739&list_no=439063&mid=a20107020000 (accessed 5 July 2026).

[2] Case, K. E. & Shiller, R. J. The efficiency of the market for single-family homes. American Economic Review 79(1), 125–137 (1989).

[3] Mankiw, N. G. & Weil, D. N. The baby boom, the baby bust, and the housing market. Regional Science and Urban Economics 19(2), 235–258 (1989). https://doi.org/10.1016/0166-0462(89)90005-7.

[4] Himmelberg, C., Mayer, C. & Sinai, T. Assessing high house prices: Bubbles, fundamentals and misperceptions. Journal of Economic Perspectives 19(4), 67–92 (2005). https://doi.org/10.1257/089533005775196769.

[5] Ministry of Land, Infrastructure and Transport. Actual Transaction Price Disclosure System. Ministry of Land, Infrastructure and Transport, Republic of Korea. https://rt.molit.go.kr/ (accessed 5 July 2026).

[6] Bank of Korea. Economic Statistics System (ECOS): exchange rates, interest rates, and monetary and liquidity aggregates. Bank of Korea. https://ecos.bok.or.kr/ (accessed 5 July 2026).

[7] Ministry of Data and Statistics. Consumer Price Survey: Consumer Price Index. Ministry of Data and Statistics, Republic of Korea. https://mods.go.kr/menu.es?mid=a20209010000 (accessed 5 July 2026).

[8] Korean Statistical Information Service. KOSIS Statistical Database: population, households, and internal migration statistics, including “Number of internal migrants for city, county, and district (M,Q,Y 1970.01–2026.05)” and ‘Netmigration(Administrative reports) (Person)’. Ministry of Data and Statistics, Republic of Korea. https://kosis.kr/eng/ (accessed 5 July 2026).

[9] Ministry of Data and Statistics. Private Education Expenditures Survey of Elementary, Middle and High School Students. Ministry of Data and Statistics, Republic of Korea. https://mods.go.kr/menu.es?mid=a20111020000 (accessed 5 July 2026).

[10] Korea Real Estate Board. National Survey of House Price Trends: housing price trend statistics and apartment housing price index. Korea Real Estate Board. https://www.reb.or.kr/rebEng/cm/cntnts/cntntsView.do?cntntsId=1372&mi=10057 (accessed 5 July 2026).

[11] Korea Education and Research Information Service. School Information Disclosure and public disclosure data services. Korea Education and Research Information Service. https://www.keris.or.kr/eng/cm/cntnts/cntntsView.do?cntntsId=1332&mi=1381 (accessed 5 July 2026).

[12] Bank of Korea. Consumer Survey Index (CSI). Economic Statistics System (ECOS), Bank of Korea. https://ecos.bok.or.kr/ (accessed 5 July 2026).

[13] Economic Policy Uncertainty. Uncertainty Indices for South Korea. https://www.policyuncertainty.com/korea_monthly.html (accessed 5 July 2026).

[14] Baker, S. R., Bloom, N. & Davis, S. J. Measuring economic policy uncertainty. Quarterly Journal of Economics 131(4), 1593–1636 (2016). https://doi.org/10.1093/qje/qjw024.

[15] Na, H., Song, W., Han, S. et al. KoTaP: A Panel Dataset for Corporate Tax Avoidance, Performance, and Governance in Korea. Scientific Data 13, 372 (2026). doi:10.1038/s41597-026-06722-5.

[16] Pires, I. M., Garcia, N. M., Zdravevski, E. et al. Daily motionless activities: A dataset with accelerometer, magnetometer, gyroscope, environment, and GPS data. Scientific Data 9, 105 (2022). doi:10.1038/s41597-022-01213-9.

[17] Tasgaonkar, P., Zade, D., Ehsan, S. et al. Indoor heat measurement data from low-income households in rural

and urban South Asia. Scientific Data 9, 285 (2022). doi:10.1038/s41597-022-01314-5.

[18] Hettiarachchi, H., Dridi, A., Gaber, M. M. et al. CODE-ACCORD: A Corpus of building regulatory data for rule generation towards automatic compliance checking. Scientific Data 12, 170 (2025). doi:10.1038/s41597-024-04320-x.

[19] Song, Y., Ahn, K., An, S. & Jang, H. Hedonic dataset of the metropolitan housing market – Cases in South Korea. Data in Brief 35, 106877 (2021). doi:10.1016/j.dib.2021.106877.

[20] Rosen, S. Hedonic prices and implicit markets: Product differentiation in pure competition. Journal of Political Economy 82, 34–55 (1974). doi:10.1086/260169.

[21] Black, S. E. Do better schools matter? Parental valuation of elementary education. The Quarterly Journal of Economics 114, 577–599 (1999). doi:10.1162/003355399556070.

[22] Bayer, P., Ferreira, F. & McMillan, R. A unified framework for measuring preferences for schools and neighborhoods. Journal of Political Economy 115, 588–638 (2007). doi:10.1086/522381.

[23] Iacoviello, M. House prices, borrowing constraints, and monetary policy in the business cycle. American Economic Review 95, 739–764 (2005). doi:10.1257/0002828054201477.

[24] Case, K. E., Quigley, J. M. & Shiller, R. J. Comparing wealth effects: The stock market versus the housing market. The B.E. Journal of Macroeconomics 5, Article 1 (2005). doi:10.2202/1534-6013.1235.

[25] Mian, A., Rao, K. & Sufi, A. Household balance sheets, consumption, and the economic slump. The Quarterly Journal of Economics 128, 1687–1726 (2013). doi:10.1093/qje/qjt020.

[26] Saiz, A. The geographic determinants of housing supply. The Quarterly Journal of Economics 125, 1253–1296 (2010). doi:10.1162/qjec.2010.125.3.1253.

[27] Gyourko, J., Mayer, C. & Sinai, T. Superstar cities. American Economic Journal: Economic Policy 5, 167–199 (2013). doi:10.1257/pol.5.4.167.

[28] Glaeser, E. L., Gyourko, J. & Saks, R. E. Why is Manhattan so expensive? Regulation and the rise in housing prices. Journal of Law and Economics 48, 331–369 (2005). doi:10.1086/429979.

[29] Sarwar, A., Bukhari, F., Sattar, A., Bukhari, M., Rehman, Z. U., Park, S. & Rho, S. Macroeconomic context-aware graph topology learning for stock price forecasting using graph neural network. ICT Express 12(1), 13–19 (2026). doi:10.1016/j.icte.2025.12.002.

[30] Bukhari, F., Sarwar, A., Sattar, A., Rehman, Z. U., Park, S. & Rho, S. Regime-aware static and dynamic graph structure learning for stock price forecasting. Humanities and Social Sciences Communications (2026). doi:10.1057/s41599-026-07653-7.

[31] Ali, Z., Ansari, Y., Bukhari, M., Maqsood, M., Park, S. & Rho, S. CMGM: A novel cross-market assets and multi-market modeling graph neural networks for financial market forecasting leveraging market states dependencies. Alexandria Engineering Journal 128, 1101–1124 (2025). doi:10.1016/j.aej.2025.08.024.

[32] Myung, S. & Kim, H. J. KRAFT: Korean Residential Apartment Fundamentals and Transactions. Version 1.0. Zenodo. https://doi.org/10.5281/zenodo.21320079 (2026).